# Alert Correlation Algorithms: A Survey and Taxonomy

Seyed Ali Mirheidari, Sajjad Arshad, Rasool Jalili

Data and Network Security Laboratory
Department of Computer Engineering, Sharif University of Technology

mirheidari@kish.sharif.edu, msarshadir@gmail.com, jalili@sharif.edu

**Abstract.** Alert correlation is a system which receives alerts from heterogeneous Intrusion Detection Systems and reduces false alerts, detects high level patterns of attacks, increases the meaning of occurred incidents, predicts the future states of attacks, and detects root cause of attacks. To reach these goals, many algorithms have been introduced in the world with many advantages and disadvantages. In this paper, we are trying to present a comprehensive survey on already proposed alert correlation algorithms. The approach of this survey is mainly focused on algorithms in correlation engines which can work in enterprise and practical networks. Having this aim in mind, many features related to accuracy, functionality, and computation power are introduced and all algorithm categories are assessed with these features. The result of this survey shows that each category of algorithms has its own strengths and an ideal correlation frameworks should be carried the strength feature of each category.

## 1   Introduction

An intrusion detection system (IDS) contains a widespread set of software or hardware whose mission is to detect improper behaviors by receiving information from their network. In terms of data processing types, such systems are divided into two categories: Anomaly-based and Misuse-based IDSs. Anomaly-based IDSs detect abnormal behaviors by checking statistical information about system execution and maintains normal behavioral patterns. Misuse-based IDSs maintain suspicious or attack patterns categories. Whenever the received information is in correspondence with the IDS signature or in contradiction with normal behavioral patterns, an alert is generated. Nowadays, many networks use such systems either commercially or open source versions. However, problems such as bad parameter settings and inappropriate IDS tuning which should be dealt with in a higher level [1]. The existence of such problems makes the alert processing system necessary. Such problems are as follows:

- **Large amount of alerts:** One of the very crucial problems of using intrusion detection systems is the large number of generated alerts by these systems. The main reasons for this large number of alerts might be imprecise incident definition, incompatibility in network, and sometimes number of real intrusions or illegal behaviors, which tend to mislead the system supervisor from the main attack or the attack goal. Anyhow, the usual number of alerts is too much to enable the system supervisor checking all of them manually. As a result, only a portion of them is checked.

- **Heterogeneous alerts:** The supervisor usually receives a wide range of alerts from different sensors and different sensors generate alerts with different formats. Hence, in order to process the alerts, it is required to normalize them.
- **False alerts and unidentified incidents:** In all types of intrusion detection systems, false in detection are made due to the lack of information describing incidents and inaccuracy of the attack pattern. Thus, a very useful activity of higher level systems is to detect mistakes made by IDSs, and correcting mistakes as much as possible. These mistakes are divided into two categories: wrong reports of illegal or unusual events which even have not occurred or their occurrence has been unsuccessful; and unreported events which must be reported.
- **Inability in connecting current alerts with the previous ones:** Nowadays, most attacks are sequential activities, which the intruders provide many phases for reaching their goals. Detecting such connections among the attack phases is sometimes very difficult, as the pattern of the first attack stage is not necessarily unique all the time and is not definitely determinable. On one hand, some attacks might not be successful due to unusual reasons or the attacker does not some parts of attack due to having direct information sources. On the other hand, the attacker might take another step of an attack in spite of the unsuccessful previous step. Being able to detect several patterns, the system supervisor would have the ability to predict the next attack step and can stop the attack before it reaches its goal.
- **Not providing the reliability level and alert priorities**: The existence of comprehensive factors for evaluating an alert importance and reliability will ease prioritizing assessments for system supervisors. But, many intrusion detection systems do not report a factor for the reliability of generated alerts, and in cases of provided criteria, the presented results are not comparable with other recourse results due to the lack of common standardization among all resources. On the other hand, the importance of an alert depends on the target importance which is not related to the IDSs. So, if a higher level system is able to assess the alert importance and priorities, it would be a valuable help to system administrators to choose alert priorities correctly.

To fulfill these requirements, Alert Correlation Systems are introduced. In the simplest manner, an alert correlation engine functions exactly as the derived meaning of the word "correlation". In fact it correlates alerts in a way that a new meaning is derived. Sometimes the number of events is so many that its manual analysis is impossible. In such cases, the correlation engine can reduce a large amount of information to a manageable rate. In addition, it can identify malicious activities from an overall and abstract view, instead of analyzing each alert separately. In other words, an alert cor-relation system is a system which receives incidents from various heterogeneous systems, reduces the required information for assessments, removes false alerts, and detects high level attack patterns.

Different algorithms have been introduced in alert correlation. To the best of our knowledge, some surveys have been presented. The first research [2] made a deep review on published papers and available tools with the aim of explaining some differences between them. In another research [3], presented a mapping among framework components and proposed techniques. In this survey, we aim to present taxonomy in which the weaknesses and strengths of previous proposed algorithms are explained. The emphasis of our survey is on some features which help correlation engine designer to propose a more accurate, practical, extendable, and low cost computation power. We believe that our approach gives a better understanding of this area as more literature works are presented and the blind area related to algorithms benefits will be kindled.

The remainder of this paper is outlined as follows. In Section 2, we provide a general categorization on alert correlation algorithms. Each category is completely introduced and its advantages and disadvantages are explained in Section 3, 4, and 5. In Section 6, we compare different algorithms based on various factors and we present conclusion and future work in Section 7.

## 2   Alert Correlation Algorithms

Alert correlation algorithms can be divided into three categories based on their characteristics: 1) Similarity-based, 2) Knowledge-based and 3) Statistical-based [4]. The similarity-based and statistical-based algorithms need less context information and they are able to correlate only based on similarities between alert features and learned information from previous steps whereas knowledge-based algorithms completely perform base on alert meanings. It has to be known that this categorization is not completely precise and some algorithms are on the edge between two categories. Thus, assigning an algorithm to a category is based on the fact that the algorithm has the most similarity to which one. Each category is introduced in the following subsections and in next sections the most important algorithms will be described.

### 2.1   Similarity-based Algorithms

The basis for this category of algorithms is defining factor to compare the similarity of either two alerts or an alert with a cluster of alerts (meta-alert). If an alert or meta-alert has needed similarity, each one of them is merged with the alert or meta-alert and otherwise a new meta-alert is created. Thus, the goal of these algorithms is to cluster similar alerts in time. The most important advantage of these algorithms is that there is no need for precise definition of attack types. Moreover, the correlation can be done only with definition of similarity factors for alerts features.

Three main subcategories are assumed for these types of algorithms. The first subcategory is based on defining very simple rules for expressing relations between alerts. The second subcategory is presented with the goal of identifying basic drawbacks in the network structure. The third subdirectory includes algorithms which produce comparison factors using models based on machine learning. In the following subsections, different researches in each subcategory will be described.

### 2.2 Knowledge-based Algorithms

This category is based on a knowledge base of attack definitions. Algorithms existing in this category are divided into two main subcategories: 1) Pre-requisites and Consequences and 2) Scenario. The basis of Pre-requisites and Consequences algorithms is on the definition of pre-requisites and possible occurring results. Thus, each incident is chained to other incidents by a network of conjunction and disjunction combinations and generates the possible network of attacks. Hence, this idea is placed in an higher level than correlation based on features similarities and in a lower level than combining based on pre-defined attack patterns. Although these algorithms do not require precise definition for each attack scenario like scenario-based algorithms, the previous knowledge is necessary for determining pre-requisites and all existing incident results. Scenario algorithms are based on the idea that many intrusions include various steps which must run one by one to success the attack. Thus, low level alerts can be compared with pre-defined intrusion steps and correlate a sequence of alerts related to each attack. Thus, a set of different attack scenarios definitions exist in a knowledge base in this type of algorithm. A list of current attack scenarios are maintained when the correlation system is operating, which this list includes all scenarios that at least one step of them are done recently. By the arrival of a new alert, it is compared to the current scenario and if the possibility is more than a certain threshold, it will attach to the scenario. Otherwise, if the alert is compatible with one of the possible scenario definitions inside the knowledge base, a new current scenario is generated using this alert. The main challenge for these algorithms is definition of attack scenarios even with existing automatic attack scenario learning methods. Also, these algorithms are completely deficient against new attacks.

### 2.3 Statistical-based Algorithms

The basic idea of these algorithms is that relevant attacks have similar statistical attributes and a proper categorization can be found by detecting these similarities. These types of algorithms store causal relationships between different incidents and analyses their occurred frequencies in the system education period using previous data statistical analysis and then attack steps are generated. After learning these relationships and being confirmed by the supervisor, this knowledge is used for correlating different attack stages. Pure statistical algorithms do not have any prior knowledge on attack scenarios. But scientific results indicate that using these algorithms is possible only in very specific domains in which domain attributes are taken in account of designing algorithms and otherwise, high error rate exist. In addition, combining data using this algorithm is impossible if the previous sensors provide incomplete or abnormal information. This category is also divided into three subcategories. The first subcatego-ry's goal is to detect alerts which are regularly repeated and finding their repetition pattern. The purpose of the second subcategory is estimating causal relationships between alerts, predicting next alert occurrence, and detecting attacks and the third subcategory's goal is combining reliability with completely similar alerts.

### 2.4 Assessment

To be able to describe the advantages and disadvantages of algorithms and assess their functionalities, we extracted several factors and explained each algorithm based on these factors. Important factors for this assessment are:

1. Algorithm Capability: Expected capabilities in algorithms are: Alert Verification, Similar Alert Clustering, Attack Sequence Detection, and Repetitive/Unimportant Alert Reduction.
2. Algorithm Accuracy: As this system is to omit incorrect alerts and combine a large number of them with the aim of expressing a summary of system states, it should have a significant exactness of errors and not to ignore any event by mistake.
3. Algorithm Computation Power: According to the high amount of calculation for correlation engine and necessity of fast and online correlation, it is necessary to assess memory usage and processing power of algorithms.
4. Required Knowledge Base: It is necessary to know the required information for each algorithm, from where this data is extracted and whether all the required data is accessible according to local presented system conditions.
5. Algorithm Extendibility and Flexibility: How much and how is the algorithm performance procedure changeable, localizable and adaptable to new conditions, by the user.

## 3 Similarity-based Algorithms

### 3.1 Simple Rules

The main idea of this subcategory can be seen in EMERALD product [5]. The functionality of this idea is based on defining very simple rules to express relations among alert features which can be combined together. In this subcategory, algorithms try to define simple rules in order to compute similarity between attributes of alerts and find the relation.

The significant works are presented in [5], [6], [7], [8], and [9]. The major required knowledge for this style of correlation rules are rule structures and required functions for checking similarity. Thus, these algorithms do not rely so much on knowledge bases. These algorithms can be used in different hierarchical levels and alerts are correlated form various aspects. In the detecting attack sequence capability, these algorithms include limits for defining attack types and can only detect sequences specified based on attack class. If the pattern definition is in a form that partitions conditions of the domain of that alert which alerts can be combined together into separate sets, it can also allocate input data to parallel processors for each pattern based on these conditions. Thus, these algorithms have a very good parallelism capability. These algorithms require maintaining all generated meta-alerts in the current time window for each pattern. Thus, its required memory is linearly proportional to alert input rate multiplied by the time window length.

### 3.2 Hierarchical Rules

This subcategory includes algorithms which have formed abstraction levels hierarchically and it makes decisions about security event detections based on these abstraction levels. This set of algorithms introduces researches that express similarity factors in a hierarchical of concept generalization.

The methods presented in [10], [11] and [12] are examples of such algorithms which is designed to detect root causes in networks. These algorithms include a method for comparing alert values, with a linear calculation degree proportional to generalization hierarchy tree depths. The required memory for these algorithms is also linear and equivalent to generalization tree sizes. Previous knowledge requirement level in these algorithms is up to defining generalization trees and thus precise and deep network structure and elements knowledge is not necessary, except in case of needing definition for address values and attack classes hierarchy.

### 3.3 Machine Learning

The last subcategory is algorithms in which comparison factors are generated automatically. Pre-requisite for supervised algorithms is the existence of a set of clustered alerts which the learning algorithm can set the parameters of its decision making model based on them. Algorithms without such a requirement (unsupervised), give the responsibility for training how to measure similarity to the algorithm. Three branches are considered for this type of algorithms.

In the first branch, the algorithms presented in [13] and [14] cluster alerts based on decision tree learning by previous data features. This algorithm exists in single-step and multi-step clustering (detecting similar alerts and attack sequences). This algorithm requires a huge and comprehensive set of training examples for creating a decision tree about how to perform correlation and in case that this set is incomplete, there is no guarantee for the correct performance of this algorithm. In terms of the required processing resources amount, it is similar to simple comparison algorithms, because each new alert must be compared to all meta-alerts existing in the current time window and the comparison procedure must be carried out with one decision tree with linear time cost. The necessary comparison structure makes it possible to divide the algorithm to few processors for comparing meta-alerts existing in the memory. Also based on the generated decision tree, partitioning input alerts and dividing them between different processors also might exist before correlation. Flexibility, compatibility with new conditions and extendibility are very hard in this algorithm and need pre-training the decision tree with new data.

In the second branch, the algorithms presented in [15] and [16] perform alert clustering based on alert Reconstruction Error by a neural network. The application of this algorithm is in single-step clustering and decision making based on cluster statistics. The only previous knowledge used in this algorithm is a set of alerts and it does not use any knowledge base and so, not using any environmental knowledge makes it hard to rely on the detection precision of this algorithm. In terms of required processing power and comparison modularity, this algorithm is very fast and simple, because of the arriving of each alert; it calculates the reconstruction error, completely independent from existing meta-alerts in the system. In case of re-learning with new condition data sets, still there is no guarantee for algorithm behavior change, because the used training model only focuses on alert re-creation precision and previous mistakes do not help the re-learning precision and new data might not have much impact on the new generalization model.

In the last branch, the algorithms described in [17] and [18] learn and apply true and false alert patterns based on labeled data by the system supervisor. The algorithm has online training capability and very good flexibility and its decision making factors are mostly based on information related to similar alerts in time ranges close to new alert arriving. Due to dependency of decision making about each alert to a wide range of similar alert statistical features, partitioning this algorithm is only possible if limiting under observation alert features to specific clusters capable of being partitioned and it is done in each unit centralized and without parallelism. This can also be mentioned for the used memory. Each executive unit in this algorithm must whether maintain statistical information related to all clusters of its own processor in the memory which is practically impossible, or check close similar cases in the permanent storage resource for each decision making. Thus, it requires a lot of access to one of the two permanent or temporary memories.

## 4 Knowledge-based Algorithms

### 4.1 Prerequisites/Consequences

The algorithms in this subcategory observe and control meanings of alerts and existing concepts in the network and then detect a security event. In addition, makes extracting and forming a relation between different attack stages possible, with the pre-assumption of knowing a knowledge base which describes all existing prerequisites/consequences of an alert, and a database describing the network configurations and structure.

One of the first algorithms in using the background knowledge has been proposed in [19]. Following this idea, a model with a simpler and applicable expression has been proposed in [20], [21], [22] and [23]. In these algorithms, alerts are modeled using first order logic and causal relationships are defined for backgrounds, and consequences of each event. Thus, a graph of possible alerts and relationships between them can be created and provide appropriate tools to reduce the amount of information shown to the user. To continue, some researches expanded the mentioned tools to identify attack scenarios, analyzed the attack procedure [24] [25] [26] and also detected lost components of an attack [27] [28].

In terms of the reliance amount on environmental knowledge, these algorithms have the most requirements and in contrast, generate conclusion outputs without any bias and completely based on real alert meanings. Given that these algorithms do not use any pre-assumed information in addition to default environmental knowledge, they are very flexible and extendable algorithms and the algorithms behavior changes in real time with any change in the environmental knowledge. In addition,

In cases of required processing power, unity, parallelism ability and required previous data, these algorithms are in the heaviest existing algorithms range, because on one hand with the arrival of each new alert, any kind of its relationship with all other alerts in the active time window must be checked and this task needs a huge amount of adjustments between alert types, prerequisites/consequences, and their information details like source and destination address. On the other hand, because these algorithms are performing around meanings, continuous maintaining and updating a lot of incidents for different resources can play a very important role in the algorithm precision. Algorithm presented in [19], [20], [21], [22], [24], [25], [26], [27], [28], [29] and [23] limit the user checking domain to overcome processing problems and making the resulting data usable. In [30] [31] [32] another method is introduced which solves the problem of requiring processing power by sacrificing the memory.

### 4.2 Scenario

The main application of this set of algorithms is detecting multi-step attacks and their reliance is on the existence scenario of these kinds of attacks. Some of these works are presented in [33], [34], [35], [36], [37], [38] and [39]. Various languages are presented for describing these scenarios but the main idea in all of them is specifying attack steps and prerequisites and its goals. Thus, in terms of required amount of environmental knowledge, this set of algorithms require a higher level of knowledge than prerequisites and results-based algorithms, but this knowledge can have less amount and domain. So, required processing resources in this branch is based on defined rules, can be less than the pre-requisites and results-based algorithms. But due to the very wide range of possible cases, unitizing and paralleling will be difficult. In case of defining a context language for expressing scenarios, these algorithms are completely flexible and extendable, because the system behavior must change in real time according to any change or extension in rules. The required memory for detecting scenarios rises according to the number of defined scenarios and required time window.

## 5 Statistical-based Algorithms

### 5.1 Statistical Traffic Estimation

In this subcategory of statistical-based algorithms, patterns of occurred alerts are recognized and the repetition pattern is derived and non-similarity with these patterns will be detected in the future.

The goal of algorithms presented in [40], [41] and [42] is creating a statistical network traffic model, predicting it, and removing predictable cases. An important category of this kind of alerts contains alerts which occur periodically, due to wrong network or security system adjustments. These algorithms do not need any context knowledge and all of their activities are carried out based on the statistics of each alert. According to this point that each of these filters is defined on a certain alert domain (according to choices made by the system supervisor), parallelism is easily possible in this application and before processing, and the alert processing unit can be easily specified. The processing load of this algorithm completely depends on the used statistical model, but all models presented in previous researches had linear and less processing load. The algorithm requirement of data set is determined based on the model time depth, but due to the use of only statistical information, storing or accessing real alerts is not necessary and only the relatively low and constant memory capacity is necessary. All presented statistical models in mentioned researches include online training and thus, compatibility with current conditions and flexibility based on new changes, are completely possible.

Also, the algorithms described in [43], [44] and [45] are expressed based on Association Rules for detecting alerts which normally occur together. An important application of this method is determining alert priorities based on this that which alerts have occurred together and have these accompaniments occurred on a usual procedure or a new pattern is observed, but also this algorithm can be used for creating related meta-alerts. Training is carried out offline in this algorithm and it is done in a time other than the execution time, but it can update model parameters in the run time and optimize the model according to new data. This algorithm does not need environmental knowledge and knowledge base and makes decision completely based on alert statistics. The algorithm requirement amount to memory is determined based on activity time window and windows are defined separately. In each window, statistical information about all alerts is calculated and the resulted statistics are compared to previous ones. Based on the determined domain by the user for applying this algorithm, arrived data from different units can be pre-partitioned and thus alerts related to each unit can be processed independently and only in their own processing unit, and processing in each unit is also very much parallelizable according to the need of counting different alert combinations.

### 5.2 Causal Relation Estimation

The purpose of this subcategory is finding alert sequence or association dominant patterns and using these patterns for detecting false cases, or proper combinations. Some of these algorithms are more proper for learning attack patterns and some for detecting false alerts or lost ones.

Several works were introduced based on analyzing causal relationships between alerts according to assessing the impact amount of using an alert in predicting other occurrence statistics of an alert [46] [47] [48]. The goal of these algorithms is creating a possible model for determining correlation relationships between alerts. Using this model, alerts can be correlated without environmental knowledge, but gaining a precise and reliable model requires a huge amount of previous data about the attacks. The algorithm acts in two separate training and functional phases. Training is performed offline and it is performable on archive data and thus its relatively huge processing load does not create any problems for practice use. In the test phase, due to the use of previous data, the model processing load is not too much and it can be decreased very much using some optimizations. Due to the high dependency of the training algorithm to huge data amount, flexing the algorithm against new conditions is not much easy, but with the direct user interference in learned relationships in the training phase, the algorithm behavior can be changed fast. Extending the algorithm is also possible using more data and wider education.

In [49], very simple algorithm for finding existing attack sequences is introduced. In this algorithm, first the possible attack graph is generated like previous algorithms and then, passed procedures in this graph are specifies in a set of previous stored alerts, and their correlation possibility is determined based on the cases that two steps of this graph have occurred in a row in real attacks [50]. In [51], an algorithm has been presented a completion of [49] idea by implementing it with Hidden Markov Model (HMM). A different feature of this algorithm is the possibility of defining the possibility of each scenario occurrence and performing each attack step based on previous steps. The knowledge of this algorithm is completely gained based on previous stored and categorized data and thus, training a strong Markov model requires a huge amount of data and correctly categorizing and specifying previous attacks. This style is focused on attacks with specified source and destinations, due to this input data are completely able of being partitioned for dividing the processing procedure into parallel processors, but the processing is relatively centered and undividable in each unit. The flexibility of this algorithm against new conditions is very slow due to high dependency to a huge amount of training data, but increasing it is possible a little by the direct interference of the system supervisor and changing Markov possibility table. For extending the algorithm, training with new labeled data is necessary.

### 5.3 Reliability Degree Combination

The goal of this subcategory is introducing an algorithm for combining reliability with completely similar alerts [52] [53]. In this type of algorithm, changing the reliability to alerts is proposed based on equivalent alert repetitions. The goal is changing the importance\priority of an alert, based on its approval by other resources. The presented algorithms require a huge amount of labeled previous data for generating probability models. The main idea can be simplified by removing all possible processing details and only accept the amount of an alert repetition as a factor independent from alert importance and resource history. The main algorithm acts in two training and function phases and thus the relatively huge amount of processing load does not have any impact on speed in time. Also the algorithm speed in execution time is completely proper and from the order of $O(1)$. Due to the independency of the algorithms process for alert clusters, input data are completely able of being partitioned and due to the high simplicity of the processing inside each cluster; there is no need of parallelism.

Flexibility against new conditions is slow due to the need of training with a lot of data, but the reliability to different resource opinions can change by the direct interference of the system supervisor and changing the algorithm behavior in real time. In addition, extending the algorithm requires extending learning data.

## 6 Comparison

In this section, we compared different algorithms from different viewpoints. In Table 1, we provided an overall comparison between three main categories of algorithms. Also, Table 2 compared all subcategories based on various factors. Considering the surveyed literature, it is obvious that in case of detection accuracy, the second category either prerequisite/consequence or scenario is high and has noticeable difference with other categories. Beside the accuracy factor, all categories have their own advantages in case of algorithm capability. We cannot ignore any of the categories because of the condition, attack and sensor type. Thus, to solve this problem, usually a hybrid approach can be used. Considering the required memory and the computation power, it should be noticed that statistical-based algorithms and the first two subcategories of similarity-based algorithms need average resources. But in the third subcategory, requirement defer according to the taken clusters. Also in the prerequisite/consequence algorithm, while there is a need for high amount of memory, there is a little need for computation power. But in contrast to prerequisite/consequence, scenario algorithms needs average resources. Another important point to be mentioned can be the weakness of statistical-based algorithm in which they have less flexibility and extendibility compared with the other categories. Also, algorithms in second category are not parallelizable because of their inner type of behavior and if partitioned, they wound have much accuracy.

**Table 1.** Overall comparison.

| Characteristic | Similarity-based | Knowledge-based | Statistical-based |
|---|---|---|---|
| Combining alerts from various sensors | Yes | Yes | No |
| Requiring Prior knowledge | Yes | Yes | No |
| Detecting false alerts | Yes | Yes | Guessing |
| Detecting multi-stage attacks | Hardly | Yes | Guessing |
| Find new attacks | Yes | No | Yes |
| Error rate | Average | Low | High |

Table 2. Comparison based on different factors

| | | Accuracy | Flexibility | Extendibility | Required Memory | Computation Power | Parallelizing | Knowledge base |
|---|---|---|---|---|---|---|---|---|
| | H — High | | | | | | | |
| | A — Average | | | | | | | |
| | L — Low | | | | | | | |
| | AD — Require Attack Definition | | | | | | | |
| | AM — Require Alert Meaning | | | | | | | |
| Similarity-based | Simple Rules | A | H | H | A | A | H | - |
| | Hierarchical Rules | A | H | H | A | A | H | AD |
| | Machine Learning (Decision Tree) | A | A | A | A | A | A | - |
| | Machine Learning (Re-creation) | A | A | A | L | H | H | - |
| | Machine Learning (Verification) | A | A | A | H | H | L | - |
| KB | Prerequisites/Consequences | H | H | H | H | L | L | AM |
| | Scenario | H | H | H | A | A | L | H |
| Statistical-based | Statistical Traffic Estimation | A | H | A | L | H | H | - |
| | Statistical Traffic Estimation (Association Rules) | A | L | L | A | A | A | - |
| | Causal Relationship Estimation (Ganger Test) | A | L | L | A | A | H | - |
| | Causal Relationship Estimation (Markov Model) | A | L | L | A | A | A | - |
| | Reliability Degree Combination | A | L | L | A | A | H | - |

## 7   Conclusion and Future Work

Regarding the analysis of many algorithms, it is necessary to take the advantages of different categories. As it is clear from the term "correlation", the more abstract the system is in networks, the better perspective the network managers have. Using more correlation measurements in this section will face great number of Events per Second (EPS). As a result, it is very important to pay attention to computation power in designing algorithms. To continue the research in future, we will take advantage of algorithms in each category to design an algorithm which has the least possible computation power consumption and can process multi thousand EPS and also has an extendable and flexible design.